# Transparent Combination of Expert and Measurement Data for Defect Prediction – An Industrial Case Study

Michael Kläs, Frank Elberzhager, Jürgen Münch
Fraunhofer Institute for Experimental Software Engineering
Fraunhofer-Platz 1
67663 Kaiserslautern, Germany

{michael.klaes, frank.elberzhager, juergen.muench}@iese.fraunhofer.de

Klaus Hartjes
Deutsche Telekom AG
Kampstraße 106
44137 Dortmund, Germany

klaus.hartjes
@telekom.de

Olaf von Graevemeyer
Deutsche Telekom AG
Hamburger Allee 25a
30161 Hannover, Germany

olaf.vongraevemeyer
@telekom.de

**ABSTRACT**
Defining strategies on how to perform quality assurance (QA) and how to control such activities is a challenging task for organizations developing or maintaining software and software-intensive systems. Planning and adjusting QA activities could benefit from accurate estimations of the expected defect content of relevant artifacts and the effectiveness of important quality assurance activities. Combining expert opinion with commonly available measurement data in a hybrid way promises to overcome the weaknesses of purely data-driven or purely expert-based estimation methods. This article presents a case study of the hybrid estimation method HyDEEP for estimating defect content and QA effectiveness in the telecommunication domain. The specific focus of this case study is the use of the method for gaining quantitative predictions. This aspect has not been empirically analyzed in previous work. Among other things, the results show that for defect content estimation, the method performs significantly better statistically than purely data-based methods, with a relative error of 0.3 on average (MMRE).

**Categories and Subject Descriptors**
D.2.9 [**Software Engineering**]: Management – *Software Quality Assurance (SQA)*.

## 1. INTRODUCTION
Planning or adjusting software quality assurance (QA) strategies in companies requires a good understanding of the expected defect content of the artifacts that undergo QA activities, as well as a good understanding of the effectiveness of the quality assurance activities themselves. Such information can support decisions about what quality assurance activities should be used, at what points in the life cycle, and to what extent.

Besides this, such information can be helpful for a variety of other purposes, such as controlling of the accurate performance of QA activities (i.e., are the expected number of defects found?). Another purpose is early risk management, e.g., by identifying cases when artifacts with high defect content are planned to undergo QA activities with predicted low effectiveness. In this case, countermeasures can be taken. Estimates about defect content and QA effectiveness estimation can also support economically-oriented decisions: Models for the economics of quality assurance describe the cost of quality, respectively non-quality, e.g. [23]. They try to provide criteria for QA trade-off decisions with respect to cost saving. Most of them require data about the expected defect content and QA effectiveness as essential input for decision-making. Unfortunately, numbers collected in empirical studies and provided in the literature are of limited use due to their great diversity [10], [1]. Estimates calculated for a specific project based on information sources from the concrete context may provide more accurate estimates.

Estimating defect content and quality assurance effectiveness is a difficult endeavor. Several approaches for these tasks have been reported in the literature, mostly focusing on purely data-based estimation or purely expert-based estimation. Applying such techniques is often challenged by the specific characteristics of the real project environments in which such techniques are to be applied. Besides the need to tailor estimation models to the specific characteristics of an organization (especially to a multitude of company-specific impact factors), other constraints such as limited availability of appropriate experts or sufficient data often exist when such techniques are piloted.

Hybrid and customizable estimation methods address some of the specific constraints of real project environments (such as limited data availability, dependence on expert judgment) and promise to assure applicability in realistic environments while providing sufficient estimation accuracy. The HyDEEP approach combines expert judgment and typically available measurement data to build prediction models for defect content and QA effectiveness [12]. The approach is inspired by other hybrid estimation methods, especially by CoBRA [2]. The HyDEEP method captures the knowledge of local domain experts in a defect content and effectiveness causal model in order to allow reuse of formalized expert experience in future projects by considering the impact of typical influencing factors in the specific context, without the need to have extensive data repositories.

HyDEEP can be used for different application scenarios in industry. HyDEEP has already been initially evaluated empirically in a specific setting in the aerospace domain [13]. The purpose of the usage of HyDEEP in that previous study was mainly *qualitative* risk assessment and QA controlling. The article at hand presents a case study focusing on another application scenario, the *quantitative* estimation of defect content and QA effectiveness with the purpose of better support planning. Unlike in the previous study, not only data about defects found by QA but also data about defects that slipped QA were available, so that quantitative estimations were applicable. The study presented here focuses on a system integration testing activity in the telecommunication domain.

The remaining paper is organized as follows: In Section 2, an overview of related work is sketched. Section 3 describes the foundations of the HyDEEP method. Section 4 presents the context and goals, the execution and analysis, and the results of the case study. Finally, Section 5 summarizes practical experience from the study.

## 2. RELATED WORK

The development and validation of prediction models to support the planning and controlling of quality assurance has been an ongoing research topic in software engineering for decades. In this section, we give an overview of existing approaches and relate them to the specific requirements and method applied in the case study.

In the area of dependable systems, one research direction is the development and application of *reliability (growth) models* [16], which use failure detection times during testing to predict the reliability of the system (and the remaining defects). These models can provide an answer to the question "When can we stop testing?", but since these models require data generated during the test process, they provide no predictions during the planning stage of the testing activity. A newer approach to predicting the defect content and effectiveness of testing activities is the use of *capture-recapture models* [21]. These models originating from biological science were successfully applied for software inspections in several studies [20]. They measure the degree of overlapping in defects found by different testers to estimate the number of defects remaining in the product. However, they can only be applied for controlling testing activities, not for planning them, because information collected during the current testing activity is required for the estimates. In addition, based on empirical results for the application of capture-recapture models in inspections, we would expect four [20] or more [24] testers have to independently test the same part of the product to obtain sufficiently accurate results.

*Fault prediction models / quality classification models* are another class of models that can be applied (at least partially) to predict the defect content of a product. They usually use a selection of metrics extracted from databases with historical product and/or process data to identify defect hotspots in the current product (e.g., most defect-prone modules) or to predict the number of defects for each module in the product. A plethora of different approaches have been developed over the last decades [4]. They mainly differentiate in two aspects: (1) the kind of measurement data they use (e.g., product-related data as measures of design complexity [3] or process-related data as the number of revisions of a module [8]) and (2) the algorithms they utilize to build the prediction model (e.g., regression or classification algorithms). Recent studies show that fault prediction models built in one context are difficult to transfer to another [17], [19]. Thus, they have to be built in each application context with a significant amount of measurement data gathered in this context. This fact inhibits their application in the presented case study, where the product is delivered as a "black box" by the subcontractor with the effect that source code or configuration management databases cannot be analyzed to obtain product- or development-process-related measurement data.

The *COQUALMO* model by Chulani and Boehm [5] – an extension of the famous COCOMO II model – can be used for project planning. It considers project cost, schedule, and quality in terms of residual defects. The model focuses on the overall development process with typical stages and QA activities, providing estimates for the defects introduced during development and the final number of residual defects. In order to support the planning of a specific QA activity, the model's abstraction level is seen as too high, especially because the set of predefined defect introduction factors may not be appropriate in specific contexts [14] and only a single factor in the model rates the effectiveness of a specific QA activity (with a five-level scale from "very low" to "extra high").

Software *process simulation models* can support decision-making by modeling and simulating the development process or parts of it [18]. In most cases, they are used to simulate the complete project; only rarely do they focus on QA or testing activities [26]. In principle, process simulation models can be built for specific QA activities based on available measurement data and expert knowledge in such a way that they can be used for defect content and effectiveness prediction. However, the fact that most simulation approaches require multiple iterations for model inspection and refinement by experienced experts as well as the use of complex and expensive tools ("Extend" and "Vensim" are the most commonly used tools [26]) may hinder their application for defect content and effectiveness prediction in practice.

In conclusion, we see that many existing approaches either concentrate on a fine-grain level – supporting the planning of tasks (e.g., fault prediction models tell us which modules to test more intensively) – or on the level of project and strategic planning, as for instance COQUALMO and many of the process simulation models. This fact limits their usefulness for QA activity-specific planning because the required data may not be available (e.g., in the case of fault prediction models) or the models are too coarse-grain to obtain precise defect content and effectiveness predictions for a specific QA activity (e.g., in the case of COQUALMO). The existing approaches that close the gap between task and project level by supporting activity-specific predictions such as capture-recapture and reliability models use data collected during the current application of the activity, which contradicts their usage for the planning of the activity.

## 3. FOUNDATIONS OF HYDEEP

The HyDEEP method is a hybrid approach for QA planning and controlling. Hybrid means that the approach is not only based on available measurement data but also tries to take advantage of the experience of available domain experts. The expert knowledge is captured in a quantified defect content and effectiveness (DCE) causal model. The quantified causal model is then used together with historical project data and a characterization of the current project to predict the defect content and effectiveness of the current QA application (Figure 1). More details on how DCE causal models are built (and applied for qualitative quality risk analysis and QA controlling) can be found in [13].

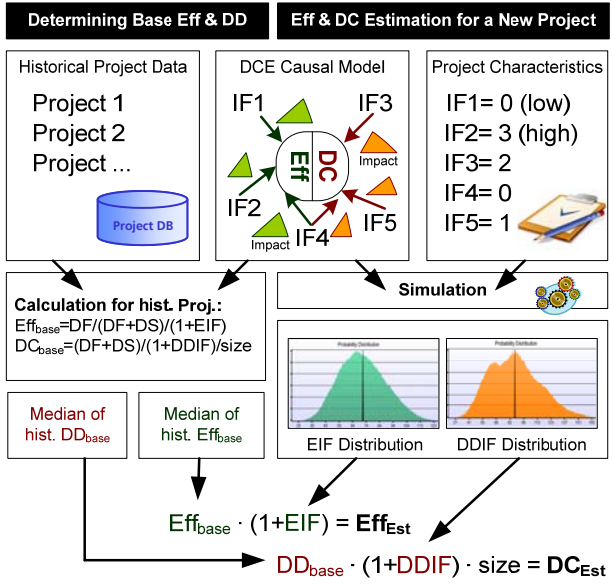

**Figure 1: Overview of the application of the HyDEEP approach for defect content and effectiveness prediction**

In the following subsections, we will focus on an overview of the principles of the HyDEEP method and show how the method can be used for predicting absolute numbers for the defect content of a product and the effectiveness of a QA activity before this activity is conducted (i.e., at its planning stage). We describe this specific kind of method application in more detail because it is the focus of the case study presented here and is not addressed in [13].

### 3.1 Defect Content & Effectiveness Equations

This subsection motivates the equations used for defect content and effectiveness prediction. *Defect density* (DD) is a common quality measure in practice and usually defined as the number of defects in a product (i.e., its *defect content* DC), divided by its size (DD=DC/size) [7]. Therefore, we can consider DD as the slope of a line in a coordinate system with product size on the x-axis and DC on the y-axis (Figure 2). When considering the DC of a product, we can assume that the product has a certain *base defect content* (i.e., the minimum number of defects in the considered context for a product of a specific size at this stage of the process) that is extended by a number of additional defects caused by factors promoting the existence of defects in the product (e.g., imprecise requirements). If we place this additional number of defects in relation to the number of defects in the best case as a relative increase (e.g., 20% more defects), we can define a relative increase factor, the *defect density increase factor* (DDIF), and provide the following equation for the defect content of a product, where $DD_{base}$ is the *base defect density* of the product:

$$DC = size \cdot DD_{base} \cdot (1+DDIF) \qquad (1)$$

The *effectiveness* (Eff) of a QA activity is typically defined as the number of *defects found* (DF) by the QA activity divided by the number of defects in the product when the QA activity started (Eff = DF/DC) [11]. This is the reason why a specific Eff value can be presented as a line in a coordinate system with DC on the x-axis and DF on the y-axis (Figure 2). Just as in equation (1) for the defect content, we can again split the number of defects found by a specific QA activity into two components. The base defects found represent the minimum number of defects found by this kind of QA activities in the considered context for a product with a given defect content. In addition to this minimum number of defects found, further defects can be detected by the QA activity when factors (e.g., the availability of experienced testers) improve the *base effectiveness* ($Eff_{base}$) of the QA activity. If we put this additional number of defects in relation to the base defects found as a relative increase (e.g., 30% more defects are detected), we can define a relative increase factor, the *effectiveness increase factor* (EIF), and obtain the following equation for the effectiveness of a specific QA activity:

$$Eff = Eff_{base} \cdot (1+EIF) \qquad (2)$$

Equations (1) and (2) can be combined into one DCE equation as presented in [13]. The DCE equation is useful if we have no information about the defect slippage (i.e., if we do not know how many defects remain in the product after the QA activity). However, using this equation does not allow predicting absolute numbers for defect content or QA effectiveness. This is the reason why we do not use the DCE equation in this case study but refine (1) and (2) based on the fact that the defect content in the product, at the start of the QA activity, is equal to the number of defects found by the activity plus the number of *defects slipped* (DS) through the activity (i.e., DC = DF + DS). The DS cannot be measured directly; instead, it is usually approximated by the number of defects found by further QA activities or in the field, which means that the DS can only be determined retrospectively for the QA activity. Applying this kind of approximation allows us to calculate the $DD_{base}$ and the $Eff_{base}$ for the QA activity for a historical project with equations (1) and (2), respectively. In order to do this, we need the defects found by the QA activity (DF), the defects found later (DS), the size of the checked product, and the project-specific DDIF and EIF (see left side of Figure 1).

### 3.2 Determine DDIF and EIF

In order to determine the context-specific values for base defect density and base effectiveness as well as to later on predict the defect content and effectiveness of the QA activity in the current project, DDIF and EIF have to be determined (Figure 3).

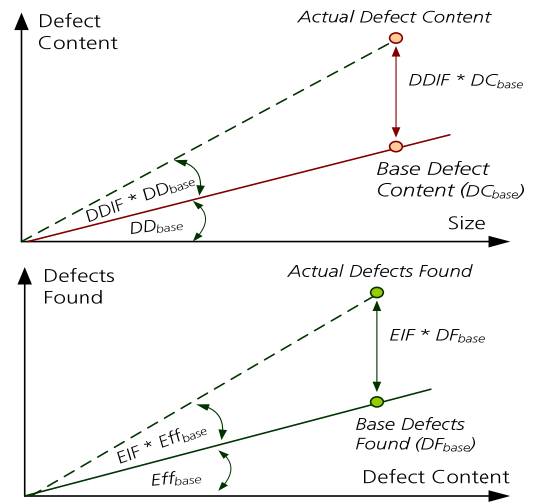

**Figure 2: Visual representation of the DC and Eff equations**

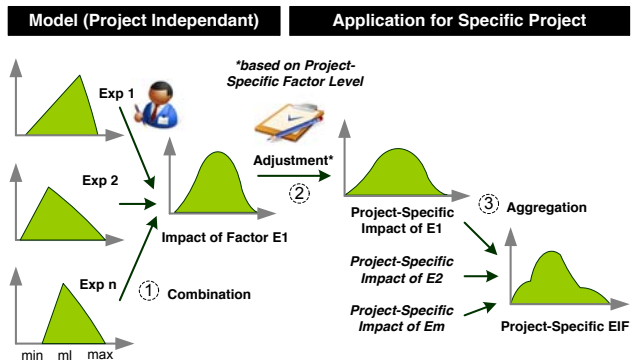

**Figure 3: Determining EIF probability distribution**

The defect density and effectiveness increase factors (DDIF and EIF) describe the project-specific increase in defect density, respectively effectiveness, relative to base values for defect density and effectiveness in the context ($DD_{base}$ and $Eff_{base}$). In order to find these relative increase factors, the most relevant influencing factors for defect density and effectiveness are captured in the DCE causal model. The impact of each of these factors is quantified by several domain experts. Comparing the context-specific best and worst cases of the respective factor, they estimate the minimum, maximum, and most likely increase in defects found (for an example, see Figure 7). The three values are then used as nodes for a triangular probability distribution capturing the uncertainty of the expert's estimate. In order to determine DDIF or EIF, (1) the triangular distributions of the different experts for one factor are combined into one probability distribution, which is (2) adjusted based on the concrete level of the factor in the considered project, and (3) finally aggregated with the adjusted impact distributions of the other defect content or effectiveness factors (Figure 3). The result of these operations is the DDIF or EIF probability distribution, respectively. The actually applied algorithms based on Monte Carlo methods are described and analyzed in more detail by [15].

## 3.3 Predicting Defect Content & Effectiveness

Assuming that we have built a DCE causal model and used it together with historical project data in the context to approximate the context-specific base defect density and base effectiveness (usually by calculating the median over the base values of the historical projects), we can predict the defect content and effectiveness for the QA activity in the current project using equations (1) and (2). The only additional information needed is the specific product size and the expert-based project characterization with respect to the factors in the DCE causal model (Figure 1).

Dependant on the base effectiveness determined for the context and the EIF probability distribution computed for a project, if we had applied the sampling algorithms used in [14] and [15] in a straightforward manner, we could have obtained a positive probability for an effectiveness value greater than 100% (i.e., predicting a certain probability of finding more defects than the product contains). If we allowed such predictions, this would contradict reality. Therefore, we have to explicitly deal with areas of the probability distribution with effectiveness values greater than 100%. We do this by restricting the value range of the probability distribution and considering areas with values greater than 100% simply as 100% (i.e., even in the case that the activity is extremely effective, we cannot find more defects than the defects that are in the product). From a probability theory point of view, this seems reasonable, because in such cases, we obtain a certain probability (>0) of finding all defects. Also, the larger the area of the probability distribution exceeding 100% effectiveness, the higher the probability of finding no defects later on.

## 4. CASE STUDY
### 4.1 Context of the Study

The presented case study took place in an integration and validation department (I&V) of T-Mobile International (TMO), where telecommunication infrastructure services are being integrated and tested. Dependant on the concrete activity, three and four members of the department, respectively, were involved in the model-building process as domain experts. All involved experts have a lot of experience in the telecommunication domain (their experience can be measured in decades). Moreover, great experience was documented concerning testing.

The product for which the prediction model was to be built is one of the main products that are validated by the department. It was released in 2004 and has been continuously maintained and extended with new features since then. The updates take place in multiple releases each year. Thus, at the time the model was built, historical data from 10 releases could be provided by TMO. The validation activity performed is a kind of acceptance test, where the system under test is provided as a black box by a subcontractor. This means that TMO has no direct access to or insights into the corresponding code. As a result, no code metrics could be measured, for example. The test process is relatively stable and uses a set of acceptance tests according to the product changes in a specific release.

### 4.2 Goals of the Study

From the scientific point of view, the primary goal of the study was to evaluate the HyDEEP method in an industrial context for the purpose of defect content and effectiveness prediction. In order to do this, a context-specific model was to be built together with available domain experts and its prediction accuracy was to be determined and compared with the accuracy of simpler models based only on the available measurement data. Moreover, questions motivated by the industry partner were investigated, too. In detail, three research questions were derived:

**RQ 1:** Is it possible to build a reasonable, context-specific, quantified causal model with acceptable effort for the local experts, which calculates effectiveness and defect content predictions of adequate quality?

This research question was split into corresponding sub-questions:

> **RQ 1.1**: How much effort is necessary for the experts during the creation phase?
>
> **RQ 1.2:** How accurate are the estimation values regarding effectiveness and defect content?

Research question one and its sub-questions resulted in the following hypotheses:

> **H$_{1.1}$:** A quantified causal model can be built with less than two person-days of effort per local expert.

**H$_{1.2}$:** The estimation error of the model with respect to defect content is significantly lower than the estimation error of applicable methods based solely on data.

**H$_{1.3}$:** The estimation error of the model with respect to effectiveness is significantly lower than the estimation error of applicable methods based solely on data.

Additionally, two more questions were investigated, which are especially valuable for motivating the application of the model in the concrete context after it was built.

**RQ 2:** Is it necessary to use each of the initially included influencing factors in the final model, and how does the prediction accuracy of the model behave when only a subset of the most relevant influence factors is considered?

**RQ 3:** How does the model behave regarding its predication accuracy when the model is initially built with a limited number of historical project data and additional project data is integrated in an iterative manner? This question is especially motivated by the practical question "How many historical releases are necessary to provide predictions with adequate accuracy?"

### 4.3 Planning of the Study

During the study, we tried to make optimal use of the most limited resource, namely the local domain experts. Therefore, we distributed all relevant steps for building a hybrid prediction model that require the involvement of the local domain experts to three workshops and three questionnaires (Figure 4). In addition, we omitted further iterations with model revisions, which are proposed in [22] for hybrid cost prediction models to improve accuracy. In order to be able to check hypothesis H$_{1.1}$, we collected the number of participating experts in each step and the effort they required.

### 4.4 Performing the Study

**Workshop 1:** The model building process started with a first workshop performed with three domain experts from TMO (i.e., experts knowing the context) and two methodology experts from the research partner (i.e., experts knowing how to apply the method). A brief overview of the method and the planned activities was presented and questions were answered.

Next, the influence factors were gathered in a brainstorming session where the experts explained what influences the number of defects found during the testing process based on their experience. All in all, eight influence factors could be identified (Figure 5). In order not to miss an important factor, a list of typical influencing factors based on [9] was used as a checklist. For each of the identified influence factors, an initial description as well as the best and worst cases for each factor in the context of the case study were defined. Two of the eight influence factors have an influence on both the effectiveness of the testing and the defect content of the tested product. Therefore, these factors were recorded twice as effectiveness and defect content factors.

Finally, the situation with respect to available measurement data for the historical releases was discussed. More details on this topic can be found under the topic "Historical project data" in this section.

**Questionnaire 1:** After the workshop, a first questionnaire was prepared by the research partner, which was used by the domain experts from TMO to rank the identified influence factors. For this, each of the five effectiveness and defect content influence factors was given a value between 1 and 5, with 1 meaning that the factor is the most important one and 5 meaning that the factor is the least important one. A factor is considered more important if the factor is responsible for more of the variation in the product's defect content, respectively in testing effectiveness in the context. The questionnaire was filled out by each of the three participants of the workshop individually and, in addition, by a fourth domain expert working in the same department.

Usually, this step is performed in order to identify the most important influence factors and include only these factors in the model. Models with too many factors (>>10) significantly increase the model building effort and may lead to instable models. In an earlier industry case, 41 influence factors were identified and ranking had to be used to identify the most important ones in order to continue building the model with an adequate number of factors [12]. Since in the TMO context, only ten influencing factors were identified, we decided during the second workshop to include all of them in the initial model. Nevertheless, the ranking was important for later analyses (RQ3). The mean and the median ranking were calculated, based on the answers from the four questionnaires.

**Workshop 2:** The factor-ranking results were presented to the experts at the second workshop. It was decided to include all identified factors in the initial causal model. No relevant interactions between the factors were identified and, therefore, each of the factors was included in the model with a direct relationship to defect content and/or effectiveness. The resulting causal model can be found in Figure 5.

Next, we defined for each influencing factor a scale with four levels (i.e., answer possibilities) to characterize a specific release with respect to the influencing factor. Using a four-level scale is not demanded by the approach, but based on good experiences in previous studies that use hybrid estimation. In this case, level 0 (low) represents the case in which the influence of the factor results in the lowest defect detection rate (i.e., defects found) and level 3 (high) represents the case in which the influence of the factor results in the highest one. The level descriptions for levels 0 and 3 should both represent realistic (i.e., observable) extremes for the given context. Rough descriptions for the best and worst cases in the context were noted in the initial workshop and were used as a basis for defining the description for the scale levels 0 and 3. Together with the experts from TMO, levels 1 and 2 were defined for each of the influence factors. One important objective was to choose the level descriptions in such a way that the impact of the factor on the defect content or effectiveness values increases linearly over the four levels as far as possible. An example of such a characterization scale can be found in Figure 6. The factor definitions and the scale level descriptions for each factor were discussed and approved.

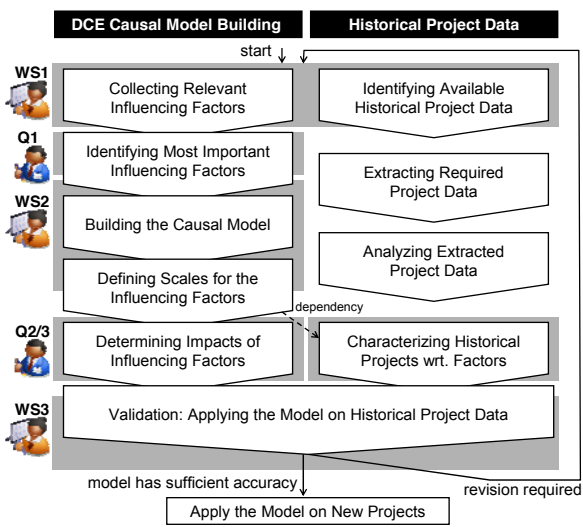

**Figure 4: Relevant model building and validation activities and their mapping to workshops and questionnaires used**

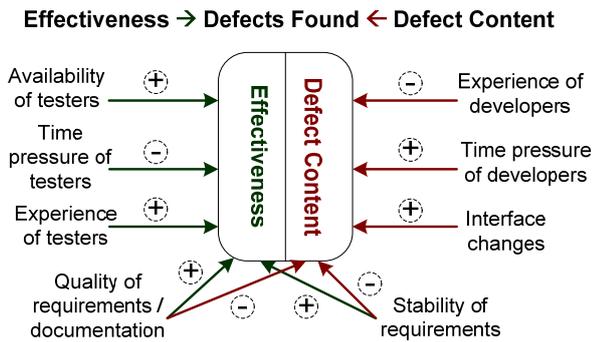

**Figure 5: DCE causal model for the TMO context**

| D3 Interface changes | |
|---|---|
| 0 | All existing requirements are easy to implement without changing interfaces |
| 1 | At least one very complex requirement with at least one new or changed interface exists |
| 2 | More than 15% of the requirements are very complex, since at least one new or changed interface exists. |
| 3 | More than 25% of the requirements are very complex, since at least one new or changed interface exists. |

**Figure 6: Example of an influence factor and its four levels**

| D3 Interface changes |
|---|
| Requirements that require changes at interfaces are typically more defect-prone, because such changes are in most cases difficult to test by the developers (they need to simulate the interface). |
| If the described factor "Interface changes" has the characteristic B instead of characteristic A, how would you rate it? |
| (A) All existing requirements are easy to implement without changing interfaces |
| (B) More than 25% of the requirements are very complex, since at least one new or changed interface exists. |
| Based on the higher defect content, I would find .. |
| F1 maximally ..  25 % more defects. |
| F2 minimally ..  10 % more defects. |
| F3 typically ..  15 % more defects. |

**Figure 7: Exemplary question for the factor quantification**

**Questionnaire 2:** The next step is to quantify the impact of each factor. Again, a questionnaire was prepared by the research partner. The experts were asked to provide estimates of the impact of the variation of the factor on the number of defects found. More precisely, they should estimate the relative increase in the number of defects found caused by increased defect density, respectively increased effectiveness, if the factor is high (level 3) instead of low (level 0). The problem with such estimations is that it is typically hard even for experts to provide estimates, especially if asked for a single precise value. A common approach to addressing this problem is to allow uncertainty in the expert's estimates by not asking for a single value, but for a minimal, a maximum, and a most likely value. These values can then be used to define a probability distribution. Based on good experience from previous studies (e.g., [13]), we used a triangular distribution. These estimates were provided by each of the four participants of the initial questionnaire. An exemplar question with fictitious values is shown in Figure 7.

**Questionnaire 3 (Expert Meeting):** Finally, a third questionnaire was prepared to characterize the historical releases regarding the influence factors. The characterization was based on the descriptions of the four levels of each influencing factor approved at the second workshop. The three experts from the initial workshop characterized the historical releases at a meeting. Although it was more time-consuming for the experts, the questionnaire was not filled out by one expert but by all three experts together in one meeting. This approach was chosen because a collaborative characterization makes the characterization more reliable and a "historically correct" characterization is critical for obtaining good prediction results. If several experts know all projects well, they can also fill out the questionnaire independently to determine their agreement.

**Historical project data:** Beside the influence factors and their quantification, the second aspect of the model when using it for prediction of effectiveness and defect content values is the data of historical projects. Klaes et al. present in [12] an overview of different application possibilities and their prerequisites. We want to apply and validate the model for quantitative QA planning (i.e., to predict numbers for defect content and QA effectiveness). Therefore, we need measurement data for the total *size* of changes in a release, the number of *defects found* by TMO in the validation activity, and the *defects slipped* through the validation.

Size: In the TMO context, information on the number of relevant and performed test cases was available as was the number of new features implemented in the release, but the features strongly vary in their complexity. Based on discussions with the experts, we finally chose the number of relevant test cases as the size variable because this value is available before a release starts and is assumed to be strongly correlated with the number of changes in a new release.

Defects found & slippage: The defects found during the validation activities and later in the field were tracked and classified by the TMO validation department. Together with the local experts, defect classes relevant for the model were identified and the number of defects found and slippage were extracted for each release. We explicitly excluded defects that were problems in the test case documentation and not in the product. In part, information about defect slippage had to be extracted from email correspondence. In summary, all relevant data were gathered for ten releases in order to build and evaluate the model.

Descriptive data analysis: Before building the model, a descriptive analysis of the data regarding effectiveness and the normalized defect content values (i.e., defect density) was performed. The goal was to analyze if outliers exist that could not be explained by typical variations between different releases. Figure 8 and Figure 9 show the results of this analysis for the ten historical projects. Due to confidentiality issues, Figure 8 and Figure 9 are provided without scales with absolute numbers.

Regarding defect density, release F was identified as an outlier, and regarding effectiveness, release H was seen as an outlier. Although G has a high defect density, the release is not considered as an outlier, because the relative increase compared to C and J is moderate when comparing it with the increase from F to E or H. The experts from TMO mentioned two possible reasons. First, the tests of the releases took place some time ago and therefore, not all relevant influences are known anymore. Second, and this was the option that they consider to be more probable, it might happen sometimes that not all found defects are documented which results in imprecise data. The total defect numbers for these releases, which are not presented here, support their assumption regarding problems with non-documented defects because for both outlier releases, the total number of defects was very low. Thus, these two releases were removed before the prediction model was built in order not to violate the validity of the model through erroneous measurement data.

## 4.5 Study Results and Model Validation

**RQ1:** Regarding RQ1, it could be stated that it was possible to gather relevant context-specific influence factors and build the causal model with them. We got positive feedback from the practitioners after discussing and defining the influence factors. Agreement could be reached regarding the overall ranking of the factors and the information was valuable.

With respect to RQ 1.1, the overall effort of the experts necessary to build the model was about one person-day. Table 1 shows the number of experts and the detailed efforts necessary for each of the activities needed to build the model. Most of the time was needed for the first workshop, where the method was explained and the relevant influence factors were gathered. The time invested at the beginning is reasonable because the identified influence factors are the basis for the model. Completing the first two questionnaires was not very time-consuming, while the joint characterization of the historical projects needed a little more time. The second and third workshops took about one, respectively two hours, to explain and discuss the results and to describe how to proceed. Besides the three experts involved in every activity, a fourth person was involved in filling out two questionnaires. Thus, hypothesis $H_{1.1}$ can be accepted.

For answering RQ 1.2, models were built based only on historical data (i.e., the influence factors were not considered) and these models were compared to the influence factor model.

<u>Prediction accuracy</u>: To compare the models in an objective way, accuracy measures proposed by Conte et al. [6] established in software estimation community were used: *relative error* (RE), *magnitude of relative error* (MRE), *mean magnitude of relative error* (MMRE), and *prediction quality* (Pred(.25)). Based on these measures, a model can be considered to be more accurate if its MMRE value is lower. In order to check the statistical significance of the improvement, a one-sided Wilcoxon Matched Pairs test was used, which is a non-parametric test and therefore a more robust counterpart of the classical paired t-test [25].

<u>Cross-validation</u>: To predict the effectiveness and the defect content values, a cross-validation approach was used; more concretely the *leave-one-out* approach. This means that to predict the defect content or effectiveness value for a specific historical release, the data from all historical releases except the one that should be predicted were used to build the prediction model. This model is then used to do the prediction. This is repeated for each of the eight historical releases used and the predicted values for defect content and effectiveness are compared with the actually measured values of these releases.

<u>Results</u>: Regarding defect content predictions, Table 2 shows the results. In the bottom part, the MRE values for the effectiveness predictions of the influence factor (IF) model are presented for the remaining eight releases, and in the upper part, the same is done for the two data-based models that can be reasonably built with the available measurement data. The first one (DC) uses simply the median of the historical defect content values for the prediction; the second one (DD) calculates the median defect density of the historical releases and uses this value together with the size data of the current release for the prediction.

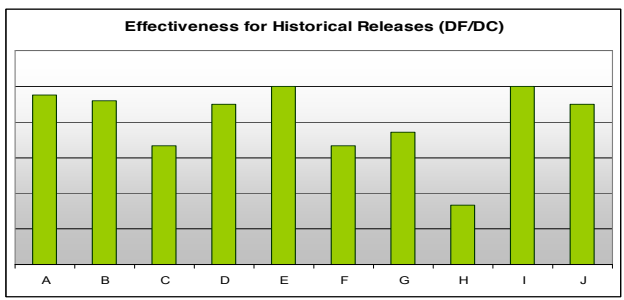

**Figure 8: Variation in effectiveness data for the ten releases**

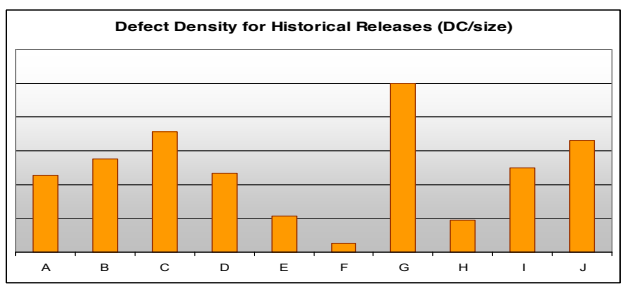

**Figure 9: Variation in defect density data for the ten releases**

**Table 1: Number of involved experts and effort needed**

| Activity | Purpose | #Experts | Effort per expert |
|---|---|---|---|
| 1st Workshop | Introduction of the method, identification of relevant factors and available data | 3 | ~3.5 h |
| 1st Questionnaire | Ranking of factors | 4 | ~20 min |
| 2nd Workshop | Discussion of ranking results, model building, introduction of the 2nd and 3rd questionnaires | 3 | ~1 h |
| 2nd Questionnaire | Quantification of factor impact | 4 | ~25 min |
| 3rd Questionnaire | Characterization of historical projects | 3 | ~1 h |
| 3rd Workshop | Presentation and discussion of model and results | 3 | ~2 h |
| Total | | 3-4 | ~ 1 day |

**Table 2: Cross-validation results for defect content prediction**

| Defect Content (DC) Model: DC = Median(DC) | | | | | | | | | | |
|---|---|---|---|---|---|---|---|---|---|---|
| Release | A | B | C | D | E | G | I | J | **MMRE** | Pred(.25) |
| MRE | 0.17 | 0.52 | 0.27 | 0.56 | 1.33 | 0.20 | 0.75 | 3.20 | **0.87** | 0.25 |

| Defect Density (DD) Model: DC = Median(DD) * size | | | | | | | | | | |
|---|---|---|---|---|---|---|---|---|---|---|
| Release | A | B | C | D | E | G | I | J | **MMRE** | Pred(.25) |
| MRE | 0.21 | 0.10 | 0.30 | 0.18 | 1.57 | 0.50 | 0.11 | 0.25 | **0.40** | 0.63 |

| Influence Factor (IF) Model: DC = size * Median (DDbase) * (1+DDII) | | | | | | | | | | |
|---|---|---|---|---|---|---|---|---|---|---|
| Release | A | B | C | D | E | G | I | J | **MMRE** | Pred(.25) |
| MRE | 0.02 | 0.00 | 0.20 | 0.23 | 1.32 | 0.45 | 0.00 | 0.21 | **0.30** | 0.75 |

**Table 3: Cross-validation results for effectiveness prediction**

| Effectiveness (E) Model: Eff = Median(Eff) | | | | | | | | | | |
|---|---|---|---|---|---|---|---|---|---|---|
| Release | A | B | C | D | E | G | I | J | **MMRE** | Pred(.25) |
| MRE | 0.05 | 0.02 | 0.38 | 0.02 | 0.10 | 0.24 | 0.10 | 0.02 | **0.12** | 0.88 |

| Influence Factor (IF) Model: Eff = Median (Ebase) * (1+EIF) | | | | | | | | | | |
|---|---|---|---|---|---|---|---|---|---|---|
| Release | A | B | C | D | E | G | I | J | **MMRE** | Pred(.25) |
| MRE | 0.02 | 0.14 | 0.35 | 0.00 | 0.10 | 0.06 | 0.10 | 0.00 | **0.10** | 0.88 |

The results of the IF model are more precise in almost every release compared to the data-based models, which is also expressed in the overall MMRE values. The predictions of release E are far from being correct and this release seems to be another outlier. Unfortunately, the reasons could not be identified, even after asking the experts from TMO. The MMRE value seems good for an initial model. Comparing the IF model with the best data-based model, the DD model, and performing a one-sided Wilcoxon Matched Pairs test with a significance level of .05, we got a p-value of 0.029, which means the improvement is significant. Thus, hypothesis $H_{1.2}$ can be confirmed.

Table 3 shows the calculated MREs regarding the effectiveness, again for the influence factor model and the data-based model. The overall MMRE is similar and very low, 0.10 on the one hand and 0.12 on the other hand. Performing a Wilcoxon Matched Pairs test with a significance level of 0.05, we got a p-value of 0.24, which means that the improvements of the influence factor model are not significant. Thus, there is no significant evidence to confirm hypothesis $H_{1.3}$. In addition, it should be mentioned that only the two most important factors influencing effectiveness (*quality of requirements documentation* and *stability of requirements*) were used in the final model because adding more effectiveness factors could not increase the prediction accuracy of the model in advance. Nevertheless, a minor improvement in the MMRE value could be observed and the IF model often presents slightly better results.

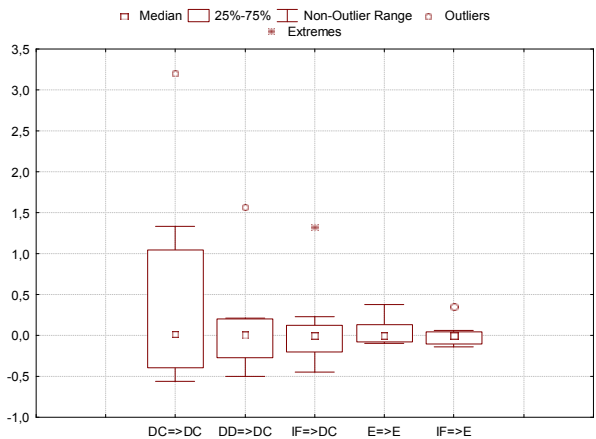

**Figure 10: Box plot with relative error of defect content and effectiveness predictions for different approaches**

Interpretation: One explanation could be that the test process in the case study context is very stable. Thus, only small influences on the effectiveness of testing activities performed can be obtained. The quality of the test case definition can be considered as the strongest influence factor on the effectiveness of the validation activity. However, most of the test cases are predefined and therefore their quality is difficult to be evaluated or influenced directly by the TMO I&V team.

RQ 2: In order to provide an answer to research question 2, different numbers of influence factors were used for the prediction. As mentioned above, for predicting effectiveness values, the final model only consisted of two influence factors. Therefore, we focused this evaluation on the five defect content factors and reduced this number step by step.

Results: Figure 11 shows the result in terms of the MMRE. The column to the far left shows the MMRE when no influence factor is used in the prediction model, which is the same as the DD model in Table 2. Thus, the value is again 0.40. Next, the factor identified as the most imported one was used in the prediction model. The overall MMRE in this case is 0.34, which is more accurate than using no influence factor. The two remaining columns show the MMRE when using the three most important influence factors and when using all five defect content influence factors (which is the original model with an MMRE of 0.30).

Interpretation: In summary, higher quality of the prediction (here expressed as a decreasing MMRE value) can be seen when using more influence factors. In the TMO context, it seems optimal to use all five defect content factors.

RQ 3: Finally, research question three asks how the model would behave over time if we were to start with a more limited set of historical releases and continuously predict new releases and then add the project data to the model.

Approach: In order to answer this question, we simulated the model building process. This means that with a low number of data from releases (and the influence factors), a prediction for the next release was done. Because we know the complete data, we could compare the predicted value with the value of the next release. Afterwards, the actual value of this release was added and the next prediction was performed and so on. From a scientific point of view, some problems with respect to validity emerge. First, a necessary minimum of releases to start with (≥4) reduces the overall number of possible increments. Second, based on a certain number of releases, one prediction for the next release is done. Based on only one value, no statistical analysis can be done. Third, by simulating the creation of the model, the real environment could not be recreated exactly. For example, the experience of the experts may change over time, which might result in different characterizations of the releases or a changed causal model. Nevertheless, this kind of analysis presents first insights into how the model would behave over time. As for RQ2, we consider only the defect content predictions, since the accuracy change in the effectiveness predictions is too small.

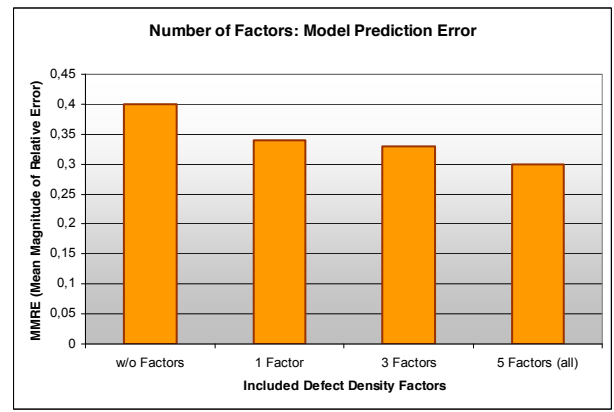

**Figure 11: Impact of different numbers of influence factors**

Results: Figure 12 shows the magnitude of relative error for defect content predictions based on different numbers of releases used. The first column shows the MRE when four releases are taken for model building (releases A-D) and release E is predicted. The relative prediction error is very high (about 1.25). When using five releases, the relative prediction error decreases to about 0.4. Starting with six releases, the prediction error seems to be very low, especially for release I and for a new release to which the model was applied (the prediction error is slightly higher for the J release).

Interpretation: Two interesting aspects could be observed. First, at least five historical releases are necessary for predictions with acceptable accuracy. Second, the quality of the prediction increases the more historical releases are available and used for the model. However, these are only initial results and their validity has to be proven in further studies.

### 4.6 Threats to Validity

As in any empirical study, there are threats to the validity of the study results [25]. Next, we will discuss what we consider to be the most relevant threats in the presented case study.

**Construct validity:** The size of the product change in each release was measured by the number of relevant test cases. This measure was chosen based on the discussion with the local experts and can be considered as a good choice considering the MMRE value for the DD model, which is based only on this information. However, we cannot validate whether this measure really represents appropriately the size of the changes in a release.

**Conclusion validity:** (i) In part, the quality of the measurement data used (especially its completeness) could not be assured. As a result, two of the historical data points (releases) had to be removed before model building, making it harder to obtain significant study results. (ii) Although we get a slightly better MMRE for effectiveness prediction with the influence factor model than with the data-based model, the statistical significance of these results could not be proven. Possible reasons may be the limited number of historical data points available, the minor effect size caused by the relatively stable testing process, or problems with the completeness of the collected defect slippage data. (iii) Due to the limited number of historical projects, we could not perform any statistical test to show the statistical significance for the result that model accuracy increases with an increased number of factors or historical releases used in the model.

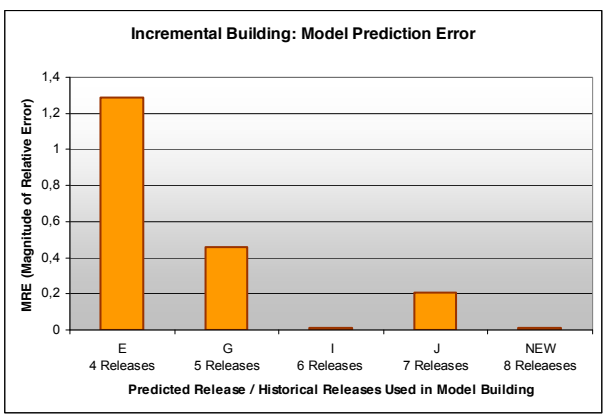

**Figure 12: Impact of different numbers of historical releases**

**Internal and external validity:** Finally, as for any industrial case study, we can see the advantage of the method's application in a realistic context, but also the influence of many context-specific factors that cannot be controlled in the study design and that may influence the results of the study, meaning that the results might not be transferable to other contexts. One such context-specific factor may be the agreement between the experts with respect to the most relevant influencing factors.

## 5. SUMMARY AND LESSONS LEARNED

In summary, we presented in this paper the first application of the HyDEEP method for the prediction of absolute defect content and effectiveness values, which can be used for the planning of a specific QA activity. The method was evaluated in an industrial case study in the telecommunication domain for the final product validation activity (system testing). In order to build the model, measurement data from historical releases were used together with expert options, which were elicited and captured in a context-specific causal model for defect content and effectiveness. In the following, we summarize the practical experience gathered in the industrial study, structured by the research questions stated in Section 4.1:

(RQ1) It was possible to build a prediction model with suitable prediction accuracy in the given context (MMRE of 0.3 for defect content and 0.1 for QA effectiveness) and with acceptable effort (3 to 4 local experts participated with an average of 1 person-day of effort for each). The defect content predictions provided by the model were significantly more accurate than the predictions by models based only on the available measurement data. A limitation could be identified regarding the effectiveness prediction in the TMO context, which could not be improved significantly compared to a purely data-based model. As the most probable explanation, we see the very stable testing process where the potentially most influential factor, the appropriateness of the largely predefined test cases for a release, could not be assessed.

(RQ2) Further, we conclude for the prediction model built in the case study that it is beneficial to include all (five) defect content factors in the model, because a reduced number of factors also reduces the model's prediction accuracy.

(RQ3) Finally, it seems reasonable to have at least five historical releases/projects to build the hybrid prediction model, which an increase in the number of available historical projects appearing to also increase the model's prediction accuracy (at least in the presented case study).

(Other) A further important lesson we learned with respect to the model building process was that the model should, if possible, contain no factors that influence both defect content and QA effectiveness. We observed that experts have great difficulty in differentiating between these two contradicting influences of a factor when quantifying its impact.

## 6. ACKNOWLEDGMENTS

We would like to thank especially the members of the T-Mobile I&V department, where we conducted the case study. We would also like to thank Adam Trendowicz and Sonnhild Namingha from Fraunhofer IESE for the initial review of the paper. This work has been partially funded by the BMBF projects TestBalance (grant 01 IS F08 D) and QuaMoCo (01 IS 08 023 C).

## 7. REFERENCES

[1] Aurum, A., Petersson, H., Wohlin, C., "State-of-the-Art: Software Inspections after 25 years", *Software Testing, Verification and Reliability*, vol. 12, pp. 131-154, 2002

[2] Briand, L. C., El Emam, K., Bomarius, F. CoBRA: A Hybrid Method for Software Cost Estimation, Benchmarking and Risk Assessment, *20th International Conference on Software Engineering*, pp. 390-399, 1998

[3] Briand, L., Wuest, J., Daly, J. W., Porter, V., "Exploring the Relationships between Design Measures and Software Quality in Object-Oriented Systems", *Journal of Systems and Software*, vol. 51, pp. 245-273, 2000

[4] Catal, C., Diri, B., "A systematic review of software fault predictions studies," *Expert Systems with Applications*, vol. 36, pp. 7346-7354, 2009

[5] Chulani S., and Boehm B., "Modeling Software Defect Introduction and Removal: COQUALMO," Technical Report USC-CSE-99-510, University of Southern California, Center for Software Engineering, 1999

[6] Conte, S. D., Dunsmore, H. E., Shen V. Y., *Software engineering Metrics and Models*, The Benjamin-Cummings Publishing Company, 1986

[7] Fenton, N. E., Pfleeger, S. L., *Software Metrics – A Rigorous & Practical Approach*, PWS Publishing Company, 1998

[8] Graves, T. L., Karr, A. F., Marron, J. S., Siy, "Predicting fault incidence using software change history," *Transactions on Software Engineering*, vol. 26, pp. 653-661, 2000

[9] Jacobs, J., Moll, J., Kusters, R., Trienekens, J., Brombacher, A., "Identification of factors that influence defect injection and detection in development of software intensive products," *Information and Software Technology*, vol.49, pp.774-789, 2007

[10] Juristo, N., Moreno, A. M., Vegas, S., "A Survey on Testing Technique Empirical Studies: How limited is our Knowledge?", *1st International Symposium on Empirical Software Engineering*, pp. 161-172, 2002

[11] Kan, S. H., *Metrics and Models in Software Quality Engineering*, Addison-Wesley Professional, 2002.

[12] Klaes, M., Elberzhager, F., Nakao, H., "Managing Software Quality through a Hybrid Defect Content and Effectiveness Model," *2nd International Symposium on Empirical Software Engineering and Measurement*, pp. 321-323, 2008

[13] Klaes, M., Nakao, H., Elberzhager, F., Muench, J., "Predicting Defect Content and Quality Assurance Effectiveness by Combining Expert Judgment and Defect Data - A Case Study," *19th International Symposium on Software Reliability Engineering*, pp. 17-26, 2008

[14] Klaes, M., Nakao, H., Elberzhager, E., Muench, J., "Support Planning and Controlling of Early Quality Assurance by Combining Expert Judgment and Defect Data - A Case Study," *Empirical Software Engineering Journal*, 2009, doi: 10.1007/s10664-009-9112-1

[15] Klaes, M., Trendowicz, A., Wickenkamp, A., Muench, J., Kikuchi, N., Ishigai, Y., "The use of simulation techniques for hybrid software cost estimation and risk analysis," *Advances in computers*, Elsevier, vol. 74, pp. 115-174, 2008

[16] Lyu, M. R., *Encyclopedia of Software Engineering*. John Wiley & Sons, chapter Software Reliability Theory, 2002

[17] Menzies, T., Greenwald, J., Frank, A., "Data Mining Static Code Attributes to Learn Defect Predictors," *Transactions on Software Engineering*, vol. 33, pp. 2-13, 2007

[18] Mueller, M., Pfahl, D., "Simulation Methods," In: *Advanced Topics in Empirical Software Engineering: A Handbook*, Springer, pp. 117-152, 2008

[19] Nagappan, N., Ball, T., Zeller, A., "Mining metrics to predict component failures," *28th International Conference on Software Engineering*, pp. 452-461, May 20-28, 2006

[20] Petersson, H., Thelin, T., Runeson, P., and Wohlin, C., "Capture-Recapture in Software Inspections after 10 years Research - Theory, Evaluation and Application," *Journal of Systems and Software*, vol. 72, pp. 249-264, 2004

[21] Scott, H. and Wohlin, C., "Capture-recapture in software unit testing: a case study," *Proceedings of the 2nd International Symposium on Empirical Software Engineering and Measurement*. ACM, New York, pp. 32-40, 2008

[22] Trendowicz ,A., Heidrich, J., Muench, J., Ishigai, Y., Yokoyama, K., Kikuchi, N., "Development of a hybrid cost estimation model in an iterative manner," *28th International Conference on Software Engineering*, pp. 331-340, 2006

[23] Wagner, S., "A model and sensitivity analysis of the quality economics of defect-detection techniques," *International Symposium on Software Testing and Analysis*, 2006

[24] Walia, G. S., Carver, J. C., Nagappan, N., "The effect of the number of inspectors on the defect estimates produced by capture-recapture models," *ACM/IEEE 30th International Conference on Software Engineering*, pp.331-340, 2008

[25] Wohlin, C., Runeson, P., Host. M., Ohlsson, M. C., Regnell, B., Wesslen, A., *Experimentation in software engineering - an introduction*, Kluwer, 2000

[26] Zhang, H., Kitchenham, B., Pfahl, D., "Reflections on 10 Years of Software Process Simulation Modeling: A Systematic Review," *International Conference on Software Process*, Springer, pp. 345-356, 2008.